\documentclass[preprint]{aastex}
\begin{document}
\title{Past and future gauge in numerical relativity}
\author{Maurice H.P.M. van Putten}
\affil{MIT 2-378, Cambridge, MA 02139-4307}
\email{mvp@schauder.mit.edu}

\begin{abstract}
  Numerical relativity describes a discrete initial value problem for 
  general relativity. A choice of gauge involves slicing space-time 
  into space-like hypersurfaces. This introduces past and future gauge 
  relative to the hypersurface of present time. Here, we propose solving 
  the discretized Einstein equations with a choice of gauge in the future 
  and a dynamical gauge in the past. The method is illustrated on a 
  polarized Gowdy wave.
\end{abstract}
\mbox{}\\
\baselineskip18pt 
 \section{Introduction}

 The initial value problem for general relativity is receiving much 
 attention in the prediction of wave-forms from candidate sources for 
 the upcoming gravitational wave detectors LIGO/VIRGO \cite{abr92,bra91}. 
 The structure of gravitational waves has recently been elucidated in new 
 hyperbolic formulations of general relativity (see, e.g., \cite{mvp01} for 
 references), which holds promise for accurate integration schemes.
 Long-time integrations also require accurate conservation of the associated 
 elliptic constraints, representing energy and momentum conservation \cite{kid01}. 
 While analytically these constraints are conserved under dynamical evolution, the 
 nonlinear nature of the equation typically tends to introduce some numerical 
 departure from exact conservation. 
 
 Gauge choice in numerical relativity involves slicing space-time into space-like 
 hypersurfaces. Data from one hypersurface are evolved numerically onto the next, 
 for example using a hyperbolic formulation. Algebraic slicing is particularly 
 illustrative, which considers prescribed lapse and shift functions $N_a=g_{ta}$ 
 for the metric tensor $g_{ab}$. 

 Here, we propose an new approach for numerical relativity with preservation
 of the constraints by through choice of gauge in the future and a dynamical 
 gauge in the past. With prescribed gauge and dynamical space-like components 
 of the metric in the future, this obtains a complete system of ten equations 
 in ten unknowns.

 The presented approach recognizes two types of constraints in Hamiltonian
 approaches within the context of the underlying four-covariant
 theory of general relativity. Recall that Hamiltonian formulations 
 assign the three-metric $h_{ij}(t)$ and its canonical momentum $\pi_{ij}(t)$ as 
 dynamical variables to a hypersurface of constant coordinate time $t$.  
 This implicitly carries a geometric constraint, in having $h_{ij}(t)$ as exact 
 projections of the four-covariant metric $g_{ab}$ onto the hypersurfaces of constant
 $t$. In addition, general relativity obeys energy-momentum conservations by the
 Bianchi identity, which translates into the Gauss-Codacci relations on the same
 hypersurfaces. These observations may be contrasted with the 
 nonlinear wave equations of \cite{mvp96}, which describe the four-covariant 
 hyperbolic evolution of the Riemann-Cartan connections $\omega_{\mu ab}$ first, 
 with subsequent slicing of space-time for the purpose of numerical implementation.
 In the continuum limit, both the projections and energy-momentum constraints are 
 exact and consistent. This need not carry over to the discrete case,
 given the nonlinear nature of the Einstein equations. This may be illustrated
 by the observed violations of energy-momentum constraints in present
 numerical simulations \cite{kid01}.
     
 A discretization of the Einstein equations introduces finite deviations from
 the continuum limit. For the purpose of numerical relativity, these deviations 
 are acceptable, provided they permit stable evolution and convergence upon 
 refinement of the discretization. The question therefore is: where are 
 discretizations allowed to introduce deviations from the continuum theory?
 The above suggests to consider seeking a trade off between the 
 geometric 
 constraint of exact projections and energy-momentum conservation. 
 This points towards a numerical
 scheme in which the past slicing is adjusted within the chosen discretization
 level, so as to satisfy energy-momentum conservation identically for a 
 given definition of the discretized Einstein tensor. 
   This obtains a complete system of evolution equations, as follows from
   simple counting: 
   the contracted Bianchi identity shows that the Einstein equations
   provide six constraints on the $2^{nd}$ time-derivatives of the
   3-metric $h_{ij}$ and four constraints on the first time-derivatives 
   of the lapse and shift functions $N_a=g_{at}$. The first 
   tells us that the discretized Einstein equations live on three
   time-slices, whereas the latter indicates that the lapse and shift functions
   are constraint on a pair of them, e.g., the first (past) and the third
   (future) time-slice. Thus, choosing a gauge for one of these
   leaves the gauge on the other determined self-consistently by
   the Einstein equations. It seems natural to maintain control
   over the future gauge, i.e., the future gauge is by choice of the
   the user. This leaves the past gauge as a dynamical variable, to
   be determined implicitly during numerical evolution. 
   This can be achieved using Newton's method.

 Non-exact projections naturally permit an uncertainty between
 the three-metric and its canonical momentum within the underlying context
 of a four-covariant theory, i.e.: also in regards to the association with 
 the hypersurface at hand. In the covariant approach of \cite{mvp96}, this 
 would thus reflect an uncertainty in the tetrad elements, which define the 
 projection, and their connections.
 This points towards a potential connection to quantum gravity. Indeed, 
 soon after this work was proposed \cite{mvp01_a}, the author learned of
 a very interesting independent discussion on the problem of consistent 
 discretizations in this context \cite{gam01}.

 \section{A discretized initial value problem}
 We illustrate our this approach on the vacuum Einstein
 equations, described by the vanishing Ricci tensor
 \begin{eqnarray}
 R_{ab}=0.
 \label{EQN_A}
 \end{eqnarray}
 The Ricci tensor $R_{ab}$ is a second-order expression in the 
 metric $g_{ab}$. Hence, (\ref{EQN_A}) defines a relationship between
 metric data $(g^{n-1}_{ab},g^n_{ab},g^{n+1}_{ab})$ on a triple of 
 time-slices $t_{n-1}<t_n<t_{n+1}$:
\begin{eqnarray}
 R_{ab}\left(g_{ab}^{n+1},g_{ab}^n,g^{n-1}_{ab} \right) = 0.
\end{eqnarray}
 Here, $R_{bd}=R^a_{\hskip0.1in bcd}$ derived from the Riemann tensor
 \begin{eqnarray}
 R^a_{\hskip0.1in bcd}=\partial_d\Gamma^a_{bc}-\partial_c\Gamma^a_{bd}
		    +\Gamma^e_{bc}\Gamma^a_{ed}
					-\Gamma^e_{bd}\Gamma^a_{ec}.
 \label{EQN_R}
 \end{eqnarray}
 This expression (\ref{EQN_R}) can be discretized by finite differencing
 on a triple of time-slices with preservation of the quasi-linear 
 second-order structure of $R_{ab}$.

 Algebraic gauge-fixing takes the form of specifying the components
 $N_a=g_{ta}$ in coordinates $\{x^a\}_{a=1}^{4}$ with $t=x^1$ time-like.
 A gauge-choice on a triple of time-slices, therefore, amounts to a 
 choice of $(N_a^{n-1},N_a^n,N_a^{n+1})$. Recall that this gauge-choice 
 in the metric arises explicitly in the Gauss-Codacci relations for 
 energy-momentum conservation. The components $h_{ij}=g_{ij}$, where 
 $i,j=2,3,4$ refer to projections of the metric into the time-slice
 $t=$const., which describe the dynamical part of the metric. The
 combination $(h_{ij},N_a)$ reflects 
 the mixed hyperbolic-elliptic structure in numerical relativity and
 (\ref{EQN_A}) represents ten evolution equations in these variables
 on a triple of time-slices. 
 
 In algebraic gauge-fixing, we prescribe $N_a^{n+1}$ as a future gauge 
 in computing $h^{n+1}_{ij}$ on a future hypersurface $t=t_{n+1}$ from 
 data at present and past hypersurfaces $t=t_{n-1}$ and $t=t_{n}$.  
 We propose closure by re-introducing $N_a^{n-1}$ as dynamical gauge
 in the past, leaving $h_{ij}^{n-1}$ fixed. Combined, this defines
 an advanced hyperbolic-retarded elliptic evolution of the metric.
 The paritioning of the metric in past and future variables as
 \begin{eqnarray}
 g_{ab}=(h_{ij}^{n+1},N_a^{n-1})=\left(
 \begin{array}{cccc}
 N_1^{n-1} & N_2^{n-1} & N_3^{n-1} & N_4^{n-1}\\
 N_2^{n-1} & h_{xx}^{n+1} & h_{xy}^{n+1} & h_{xz}^{n+1}\\
 N_3^{n-1} & h_{xy}^{n+1} & h_{yy}^{n+1} & h_{yz}^{n+1}\\
 N_4^{n-1} & h_{xz}^{n+1} & h_{zy}^{n+1} & h_{zz}^{n+1}
 \end{array}
 \right)
 \label{EQN_PG}
 \end{eqnarray}
 thus obtains ten dynamical variables in the ten equations
 \begin{eqnarray}
 R_{ab}(h_{ij}^{n+1},N_a^{n-1},\cdots)=0~~\mbox{at}~~t=t_n.
 \label{EQN_B}
 \end{eqnarray}
 Here the dots refer to the remaining data
 $(h_{ij}^{n-1},h_{ij}^n,N_a^n,N_a^{n+1})$, which are kept fixed 
 while solving for $(h_{ij}^{n+1},N_a^{n-1})$.
 Thus, (\ref{EQN_B}) which takes into account {\em all} ten Einstein
 equations with no reduction of variables. Time-stepping by (\ref{EQN_B}) 
 {\em evolves the metric into the future with dynamical gauge in the
 past}, in an effort to satisfy energy-momentum conservation within the
 definition of the discretized Ricci tensor.
Because (\ref{EQN_B}) comprises derivatives of $N_a$ only to first-order 
in time, numerically though the data $N^{n+1}_a$ and $N^{n-1}_a$,
we anticipate that the evolution of $N_a$ is of first-order 
in the $t-$discretization $\Delta t$. This introduces
non-exactness in $h_{ij}^{n-1}$ as projections of $g_{ab}$ on 
$t-\Delta t$ to within the same order of accuracy. It may result in a 
first-order drift in the $t-$labeling of the hypersurfaces -- 
permitted by coordinate invariance.
 
 The presented approach can be illustrated on a polarized Gowdy wave.
 Gowdy cosmologies are an extensively studied class of universes with 
 compact space-like hypersurfaces with two Killing vectors 
 $\partial_\sigma$ and $\partial_\delta$. With cyclic boundary conditions, 
 the space-like hypersurfaces are homeomorphic to the three-torus as a 
 model universe collapsing into a singularity. The associated line-element 
 is (see, e.g., \cite{ber93})
 \begin{eqnarray}
 ds^2=e^{(\tau-\lambda)/2}\left(-e^{-2\tau}d\tau^2+d\theta^2\right)
      +d\Sigma^2,
\label{EQN_G1}
 \end{eqnarray}
 where $\lambda=\lambda(\tau,\theta)$ and $d\Sigma$ denotes the surface 
 element in the space supported by the two Killing vectors. Polarized 
 Gowdy waves form a special case, which permit a reduction to
 \begin{eqnarray}
 d\Sigma^2=e^{-\tau}\left(e^Pd\sigma^2+e^{-P}d\delta^2\right).
 \label{EQN_G2}
 \end{eqnarray}
 Here $P$ satisfies a linear wave-equation
 $P_{\tau\tau}=e^{-2\tau}P_{\theta\theta}$; a long wave-length solution is
 \begin{eqnarray}
 P_0(\tau,\theta)=Y_0(e^{-\tau})\cos\theta,
 \label{EQN_G3}
 \end{eqnarray}
 where $Y_0$ is the Bessel function of the second kind of order zero.
 This leaves
 \begin{eqnarray}
 \lambda(\tau,\theta)=\frac{1}{2}Y_0(e^{-\tau})Y_1(e^{-\tau})
        e^{-\tau}\cos2\theta
       +\frac{1}{2}\int^1_{e^{-\tau}}\left(Y_0^{\prime 2}(s)+Y_0^2(s)\right)
	   sds.
\label{EQN_G4}
\end{eqnarray}
A spectrally accurate numerical integration is described in
\cite{mvp97}.

The implicit equation (\ref{EQN_B}) for the dynamical variables 
$(h^{n+1}_{ij},N_a^{n-1})$ has been implemented numerically. We have 
done so by solving for the all ten components 
$(h^{n+1}_{ij},N_a^{n-1}$)
using Newton iterations on these variables. This procudere uses a
numerical evaluation of the Jacobian
\begin{eqnarray}
J_{AB}=\frac{\partial R_{A}}{\partial U_B} 
\end{eqnarray}
where the capital indices $A,B=1,2,\cdots,10$ refer to the labeling
\begin{eqnarray}
\begin{array}{rl}
R_A& =(R_{11},R_{22},R_{33},R_{44},R_{12},R_{13},R_{14},R_{23},R_{24},R_{34}),\\
U_B& =(N_1^{n-1},h_{11}^{n+1},h_{22}^{n+1},h_{33}^{n+1},N_2^{n-1},N_3^{n-1},
       N_4^{n-1},h_{23}^{n+1},h_{24}^{n+1},h_{34}^{n+1}).
\end{array}
\end{eqnarray}
The Ricci tensor (\ref{EQN_R}) has been implemented by second-order
finite differencing, such that it remains quasi-linear in the second
derivatives. 
In particular, the Christoffel
symbols
\begin{eqnarray}
\Gamma_{ab}^c = \frac{1}{2}g^{ce}(g_{cb,a}+g_{ac,b}-g_{ab,c})
\end{eqnarray}
is obtained by symmetric finite-differencing on the metric
components, and itself differentiated by the product rule following
individual numerical differentiations of $g^{ab}$ and 
$(g_{cb,a}+g_{ac,b}-g_{ab,c})$.
The choice of future gauge $N_a^{n+1}$ is provided by the 
the components
\begin{eqnarray}
g_{at}=(e^{(\tau-\lambda)/2},0,0.0)
\end{eqnarray}
of the analytical line-element 
(\ref{EQN_G1}-\ref{EQN_G4}), which facilitates error analysis
by direct comparison of the numerical results with the analytic 
expression for the line-element (\ref{EQN_G1}). It will be appreciated
that in principle other choices of $N_a^{n+1}$ can be made. 

Fig. 1 shows numerical results for evolution of initial data
on $0\le\tau\le 4$. The results show that {\em all} Einstein
equations in the form of $R_{ab}=0$ are satisfied with arbitrary
precision, while the metric components are solved accurately to 
within one percent. The asymptotic behavior of the implicit
corrections to the lapse functions are shown in Fig. 2. Note that
these corrections are finite to first-order in $\Delta t$, 
a testimonial to the appearance of the lapse function
in the Einstein equations to first-order in time.

In summary, a dynamical gauge in the past gives a complete number of 
ten degrees of freedom in evolving to a future hypersurface, permitting
{\em all} ten Einstein equations to be satisfied with arbitrary
precision. The Einstein equations are hereby satisfied numerically on a 
triple of hypersurfaces of past, present and future within the definition 
of the discretized Riemann tensor.
The simulation of a nonlinear one-dimensional Gowdy wave by 
implicit time-stepping according to the ten discretized vacuum Einstein 
equations (\ref{EQN_B}) serves to illustrate a numerical implementation. 
The presented approach ensures conservation of energy and momentum,
within the definition of the discretization used. Satisfying these constraints
is generally a necessary condition 
for stability in long-time 
integrations. It is an important open question if satisfying these constraints
is a sufficient condition for stability.
It would be of interest to study this method in evolving
Schwarzschild black holes with singularity-avoiding slicings of spacetime,
and its behavior in the presence of outgoing radiative boundary conditions.
More generally, it would be of interest to consider this approach in higher 
dimensions, including a self-consistent integration of any of the modern 
hyperbolic formulations and efficient elliptic solvers.

{\bf Acknowledgements.} This research is partially supported by
NASA Grant 5-7012 and an MIT C.E. Reed Award. The author thanks
P. Hoefflich for drawing attention to the asymptotic behavior of
slicing and L. Wen for stimulating discussions.

\mbox{}\\
\mbox{}\\
\mbox{}\\
\centerline{{\bf Figure Captions}}

\mbox{}\\
{ {\bf FIGURE 1.} Shown is the simulation for $0\le\tau\le 4$ of
the polarized Gowdy wave. The solutions $P(\tau,\theta)$ and 
$\lambda(\tau,\theta)$ are displayed as a function of 
($\tau,\theta)$ (upper windows).
The middle windows display the solutions for $\tau=4$,
wherein the circles denote the numerical solution and the
solid lines the analytical solution.
The $\tau-$evolution of the errors (lower windows) 
are computed relative
to the analytical solution to Gowdy's reduced wave equation. 
The simulations discretize $\theta$ by $m_1=64$ points and 
the $\tau-$interval by $m_2=1024$ time-steps. 
Particular to the proposed numerical algorithm is a dynamical gauge
in the past and a prescribed gauge in the future. This permits
satisfying {\em all} of the discretized Einstein equations 
$R_{ab}=0$ to within arbitrary precision by Newton iterations.
The slight increase in the error of about $10^{-10}$ reflects
the exponential growth of the analytic solution, because the
Gowdy cosmology evolves towards a singularity.

\mbox{}\\
{\bf FIGURE 2.} Shown are the self-consistent corrections on the
slicing $t=t_{n+1}$, introduced by the difference between the past gauge 
$N^{n-1}(t_{n+2})$ to the hypersurface $t=t_{n+2}$ and the earlier future gauge 
$N^{n+1}(t_n)$ to the hypersurface $t=t_{n}$. The three curves refer
to different discretizations $m_1=16,32$ and $64$ points with, respectively,
$m_2=256,512$ and 1024 time-steps. These similar results for various
discretizations indicate asymptotic
behavior consistent with the first-order appearance of the lapse function 
in the Einstein equations. A first-order accuracy in lapse introduces a
commensurate offset in slicing or, equivalently, an offset in the coordinate $t$.


\newpage

\plotone{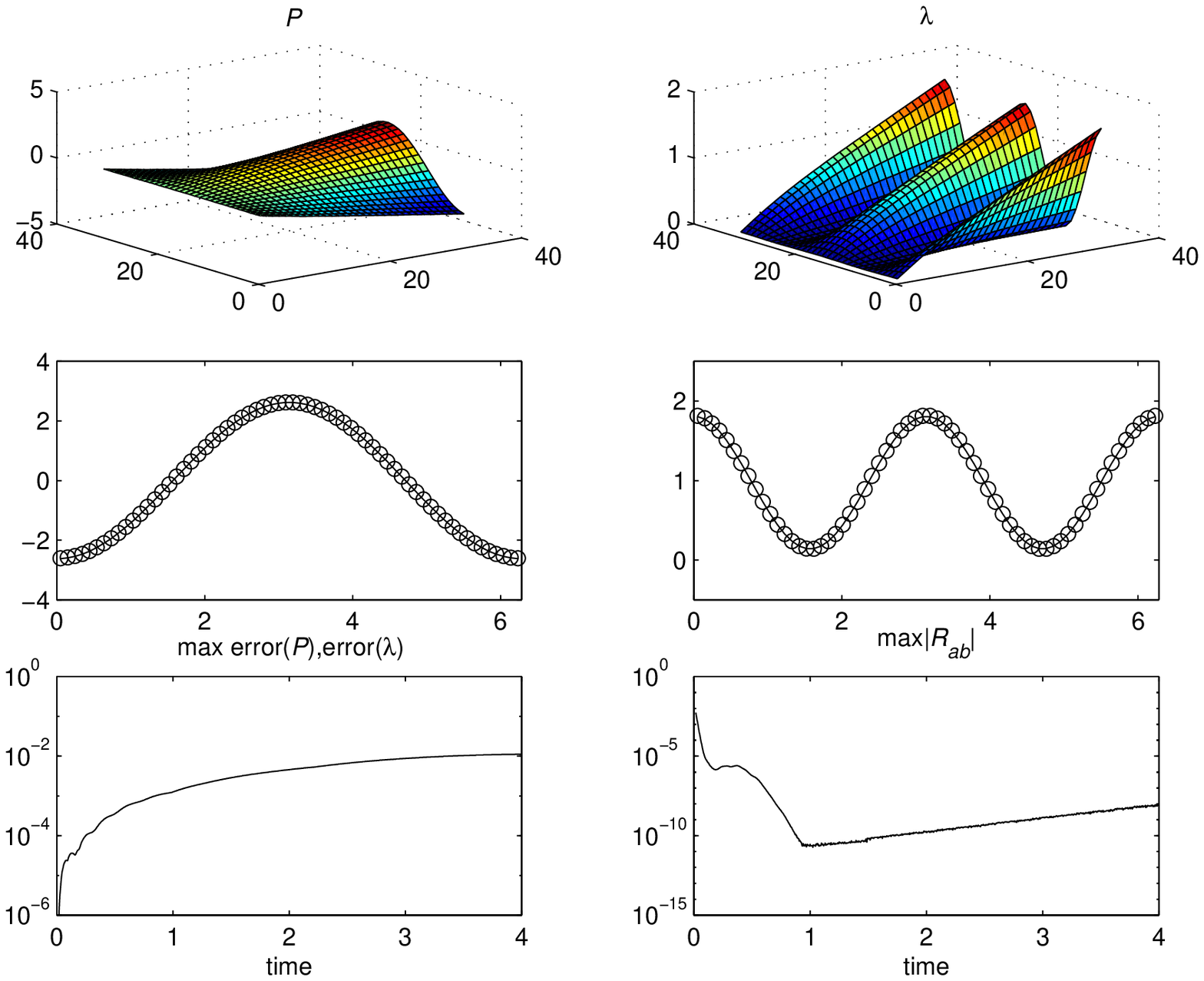}
\vskip2in {\sc FIGURE 1}

\plotone{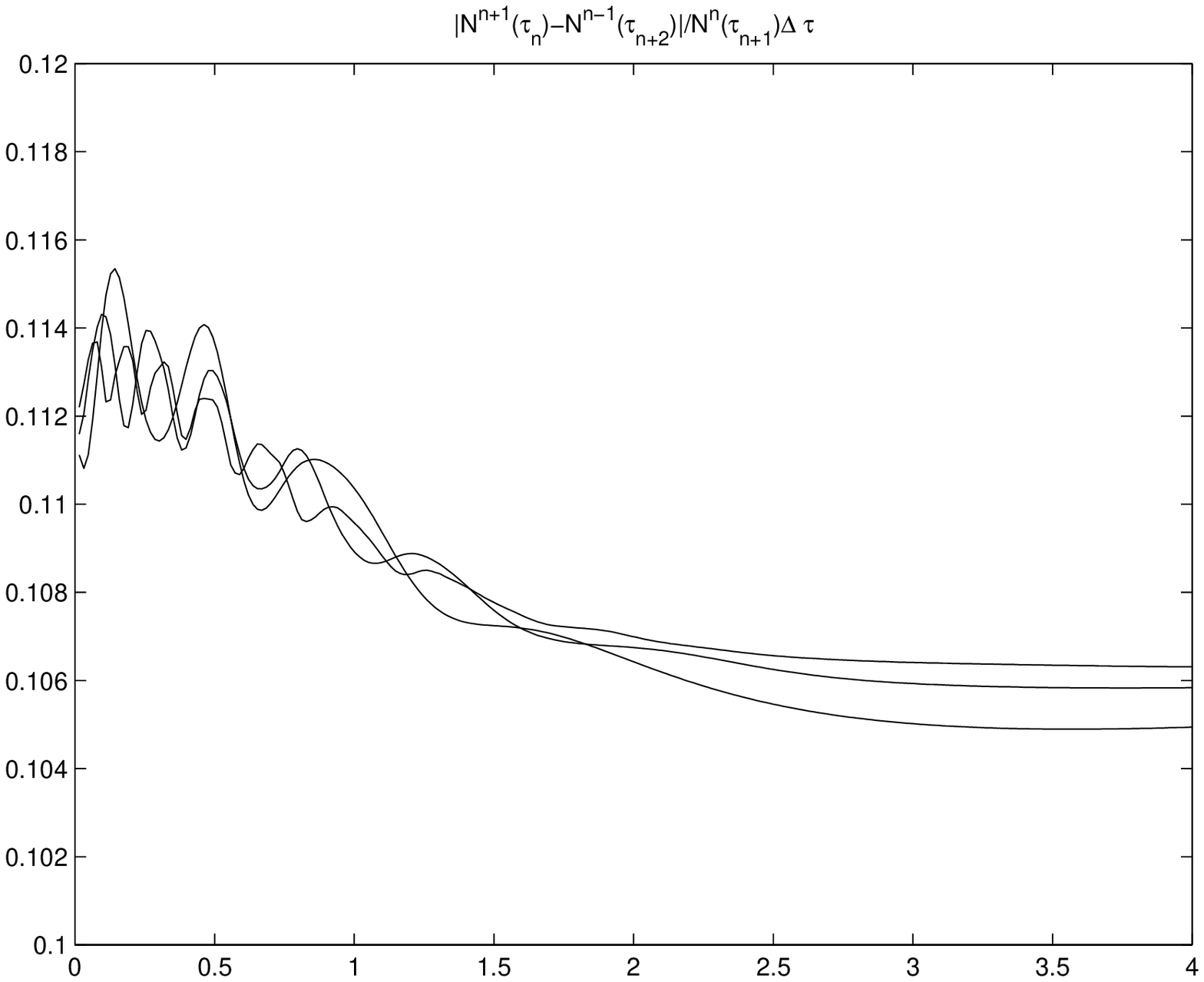}
\vskip1in {\sc FIGURE 2}

\end{document}